\documentclass[11pt]{article}
\usepackage{amsmath,amsfonts,amssymb}
\usepackage{bm}
\usepackage[utf8]{inputenc}
\usepackage[english]{babel}
\usepackage {eucal}
\usepackage{graphicx}
\usepackage{color}
\usepackage{float}
\usepackage{arydshln}
\usepackage{mathtools}
\newcommand{\llra}{\leftrightarrow}
\newcommand{\ra}{\rightarrow}

\newcommand{\wh}{\widehat}

\newcommand{\be}{\begin{equation}}
\newcommand{\ee}{\end{equation}}
\newcommand{\bea}{\begin{eqnarray}}
\newcommand{\eea}{\end{eqnarray}}

\begin{document}

\pagestyle{empty}

\begin{center}
\vspace{5cm}
%\begin{frontmatter}

{\Large \textbf{Hierarchy of codon usage frequencies  from  \\ \vspace{4mm} codon-anticodon interaction in the crystal basis model }}
%\textcolor{red}
\bigskip

\bigskip

\bigskip

\bigskip
%\end{center}

\vspace{8mm}

{\large   A. Sciarrino}

\vspace{4mm}

 \emph{Former Professor \\ University of Naples  ``Federico II" \\ nino.sciarrino@gmail.com}
 
 \vspace{8mm}

{\large   P.Sorba}

\vspace{4mm}

 \emph{LAPTH,Laboratoire d'Annecy-le-Vieux de Physique Th\' eorique CNRS \\ Universit\' e de Savoie Mont Blanc\\
 Chemin de Bellevue, BP 110,\\
 F-74941 Annecy-le-Vieux, France \\
E-mail: paul.sorba@lapth.cnrs.fr}
\end{center}
\vspace{10mm}

 \vspace{8mm}
\begin{abstract}
Analyzing the codon usage frequencies of a specimen of 20 plants, for which the codon-anticodon pattern is known, we have remarked that the hierarchy of the usage frequencies present an almost ``universal"  behavior. Searching to explain this behavior,   we assume that the codon usage probability results from  the sum of two contributions: the first dominant term is an almost ``universal" one and it depends  on the codon-anticodon interaction; the second term is a local one, i.e. depends on the biological species.  The codon-anticodon interaction is written as a spin-spin plus a z-spin term in the formalism of the crystal basis model. From general considerations, in particular from the choice of the signs and some constraints on the parameters  defining the interaction, we are able to explain most of the observed data. 
\end{abstract}
\vspace{8mm}
\leftline{LAPTH-002/23}
%\leftline{January 2023}
\vspace{8mm}

\newpage

\pagestyle{plain}
\setcounter{page}{1}

\section{Introduction}

The study of the  pattern codon-anticodon is very complicated as there is no one to one correspondence between codon and anticodon. 
One anticodon can recognize and bind to more than one codon, through the ``wobble mechanism”.
There are 61 coding codons and at most 46 anticodons for the eukaryote code \cite{MG02}, which are used with different frequencies, depending on the biological species and even on the particular genes inside the organism. 
The non-uniform usage of codons is referred as  ``codon bias".
The ``codon bias" has been and is object of a countless number of articles addressing the study of its causative factors, of its dependence on GC content and on the  genes expression level, of relative abundance of a specified codon in different genes of the same biological species or in the same gene for different biological species. Little
attention has been paid to analyze the codon usage frequency averaged over the sequences of many biological species to infer possible global correlations between them.

 We and other coworkers have studied this aspect mainly focusing on the structure of the quadruples, i.e. the four codons with the same first two nucleotides (di-nucleotide).
Indeed,  in \cite{FSS03}, for vertebrates species , it has been derived that the sum of the codon usage frequencies for codons belonging to each quartet, i.e. the  set of four codons  differing only for the third nucleotide and encoding the same amino-acid with third nucleotide C and A (or G and U) is a constant (sum rule), that is independent of the considered biological species  but depends on the quartet and on the GC content, and in \cite{FSS05}  investigation of the statistical reliability of the observed pattern has been carried out. Note that we consider eight quartets, as we consider the three sextets as the sum of a quartet and of a doublet. In \cite{CS16} this result  has been extended to species belonging to invertebrates and plants. In  \cite{AG07}  the investigation has been extended also to fungi and bacteria. Moreover it has been remarked that the constant of the sum of the codon usage frequency ending with C and A (or U and G) holds for all quadruples, not only for the quartets which encode the same amino acid, in general better than any other couples of nucleotides. This property holds with a much lower degree of accuracy, in particular for the two quadruples which include Stop codon, as expected. These previous studies suggest that the structure of a quadruple, for the codon usage frequencies, may be more relevant than usually considered.  Consistently, we will analyze the data of the published literature focusing on the quadruple aspect and not, as usually, on the sets of codons coding the same amino acid.

In \cite{Salvia}-\cite{Ulva} 
 the entire genome   
of different plants has been analyzed  and  the codon usage frequencies and the codon-anticodon pattern, i.e.  which anticodon recognizes which codons, of these biological species has been reported.
 In the following we analyze the data reported in \cite{Salvia}-\cite{Glyc}, splitting or regrouping them in quadruple.
We do not report the data for biological species  belonging to the same family because, as expected, they are very similar. 

From the data reported in \cite{Salvia}-\cite{Ulva}, we compute the codon usage frequency, see Tables \ref{Table:Sal-Coc}-\ref{Table:Ulva} of Appendix 1. 
 The data reported in Table \ref{Table:Ulva} will not be included in our discussion and will be commented separately. In fact, while the GC content of the coding sequences of the first 20 species  ranges between about 37 \% and 40 \%, the GC content of {\it Ulva flexuosa} reported in  \cite{Ulva} is about 27 \%, so much lower than the previous ones. In fact, from comparison of the frequencies reported in Table \ref{Table:Ulva} with the frequencies reported in the other Tables, one can deduce that GC content difference is a very determining factor.

 Looking at these Tables  we remark the following pattern: 
 \begin{itemize}
\item  the most used codon in any quadruple is the one ending with U or A.

\item if the most used codon in a quadruple ends with U, then the most used codon in the quadruple corresponding to the Watson-Crick or  conjugate di-nucleotide,  ends with A,  except  the couples CGN-GCN, GUN-CAN  (N = C,U,G,A).  Let us recall that the conjugate nucleotides or canonical base-pairs are: $ C \llra G, U \llra A$. Moreover the rank or hierarchy of the codon usage frequencies generally exhibits a conjugate or  ``mirror"  pattern which will be further detailed after introducing our notation.
 \end{itemize}

The remarked pattern is almost ``universal", i.e. it is  the same for almost all the plants of the considered specimen. 

The fact that  the most used codons end with U and A has been already observed in the literature and it can be understood as the analyzed biological species have an abundance of UA. The others features of the rank of the codon usage frequencies do not seem to be already observed neither can  be explained  by the UA abundance.  The generality of the observed pattern of the codon usage frequencies, even if our statistics is rather limited, suggests that its origin is not casual, but must have a general underlying reason.
 
 The aim of the following study is to try to explain these general features of  the codon usage frequencies from the codon-anticodon interaction in the crystal basis model  \cite{FSS98}. For a detailed review of the model with its applications, see  \cite{SS17}.

Indeed we shall find that a possible explanation of the observed pattern can be obtained 
in the framework of the cited model from the interaction  introduced in \cite{SS12} and applied for determining the minimum number of anticodons  for animal mitochondrial code.

 In  \cite{SS16},  from the minimization of this codon-anticodon interaction and from some very general assumptions, we have obtained semi-quantitative inequalities for the codon usage bias for different sets of biological species. The outcomes derived are in an amazing agreement with the observed data computed by averaging  the values of the codon usage frequencies, $P_N$, see eq.(\ref{eq:1}) in the next Section for its precise definition, over a specimen of different biological species. The agreement between the theoretical prediction and the observed data, despite the over-simplifying assumptions of our theoretical scheme, suggests that our approach may capture some relevant features of the underlying relevant mechanism, indicating, as already accepted in the literature, that the codon–anticodon interaction plays a relevant role in the codon bias. 
 In particular, in \cite{SS16} we made the simplifying assumption the codon usage frequencies were \emph{a priori} fixed quantities, while in the present paper we shall determine not  their values, but the hierarchy of their values, the guiding line in our approach, also in the present paper, being the minimization of the  codon-anticodon interaction. 

The paper is organized as it follows: in Section 2 we introduce our notation, in Section 3 we discuss our model  in the framework of Watson-Crick codon-anticodon recognition pattern, in Section 4 the model is applied to the real situation, in Section 5 we discuss the results and we make the comparison with the experimental  data. Finally a few conclusions and highlights are presented in the final Section. In Appendix 1 tables with codon usage frequency for the biological species analyzed are reported.  In Appendix 2 a very short summary of the model and of the notations is given. 

\section{Notation for the quadruples}

In the following we use the standard notation for the nucleotides: $N = A,C, G, U; Y = C,U; R = G,A$.

 Let us define the codon usage frequency of a codon in a quadruple,
\begin{equation}
\label{eq:1}
P(XZN) = \frac{n_{XZN}}{n_{XZ}}
\end{equation}
where $n_{XZN}$ ($n_{XZ}$) is the observed number of times the codon $XZN$ (respectively  the observed total number of the codons in the quadruple, $n_{XZ} = \sum_Nn_{XZN}$) appears in the protein biosynthesis.
To simplify the notation, in the following we denote, for any fixed di-nucleotide $XZ$, the frequency  $P_N \equiv P(XZN)$, normalized 
\be
\sum_{ N = A,C, G, U} \; P_N = 1 \label{eq:n}
\ee

We assume that, considering entire genomes, the codon usage frequency can be assimilated to the theoretical probability
\begin{equation}
\label{eq:1-t}
P(XZN)_{th} =  \lim_{n_{XZ} \to \infty} \;\;\; \frac{n_{XZN}}{n_{XZ}}
\end{equation}

In the following we assume that:
\medskip

{\bf Assumtion 1 -  The probability distribution  $P(XZN)_{th}$ for any codon, except the Stop codons, can be written  as the sum of two contributions. The first one, denoted by $\wh{P}(XZN) \equiv \wh{P}_N$, is a ``universal" effect, in the sense that it depends on the  codon-anticodon interaction and, for any fixed di-nucleotide $XZ$,  ``smoothly" depends on the biological species, in a suitably defined biological domain with almost the same GC content range.  The second one, denoted by $f(XZN) \equiv \rho_N$, is a  ``random” or ``local" effect depending on the biological species. In the following we will specify what we mean by ``smooth" dependence.}

\medskip

So, for any  codon quadruple we write:
\be
P(XZN)_{th} =  \wh{P}(XZN) +  \rho(XZN)  \equiv \wh{P}_N  + \rho_N \approx  P(XZN)  \equiv  P_N
 \label{eq:P}
\ee
we assume the normalization:
\be
\sum_N \; \wh{P}_N = 1
\label{eq:norm}
\ee

The normalizations eq.(\ref{eq:n}) and eq.(\ref{eq:norm})  imply
\be
\sum_N \; \rho_N = 0
\label{eq:norm0}
\ee
Let us remark that the parametrization used in eq.(\ref{eq:P}) has been introduced for quartets in \cite{FSS03}. In the present  paper, we extend it to all the quadruples.

We shall analyze fourteen quadruples of codons, not considering the ones containing one or two Stop codon (i.e. UGN and UAN). 

The basis of our analysis is the following:

\medskip

{\noindent 
\bf Assumption 2 - In eq.(\ref{eq:P}) the term $\rho_N$ is generally smaller than $\wh{P}_N$. $\wh{P}_N$ is determined from a function of the codon-anticodon interaction. This function, at present undetermined, is assumed to be decreasing in the value of the interaction. Therefore the rank or  hierarchy of the values  of  $\wh{P}_N$ determines  the hierarchy of values of $P_N$. It follows that the most used codon in the quadruple XZN is the one with the largest value of  $\wh{P}_{XZN}$, i.e. with the smallest value of the interaction, and so on for all the rank.} 

\medskip
 
Consequently the effect of $\rho_N$ is to change the value of  $P_N$ with respect to $\wh{P}_N$ without effecting generally its position in the rank of the values. Of course we expect this statement to be better verified for the higher values of $\wh{P}_N$.
 
 \section{The model for the Watson-Crick codon-anticodon recognition pattern}

In order to get a first insight, we focus on a simple model, without the wobble hypothesis, that we call the Watson-Crick codon-anticodon recognition pattern.  That is we consider the model where any codon is read or recognized by the anticodon which is its conjugate, namely  the anticodon made from the canonically paired bases. Consequently to our assumption, the codon $X^cZ^cN^c$ is recognized by the anticodon $N^a_cZ^a_cX^a_c$, where the subscript $c$ denotes the canonically paired nucleotide and the superscript $c$ ($a$) denotes the nucleotide belonging to a codon (respectively to an anticodon). We shall explicitly  use this notation whenever a confusion may show up.  Note that we write both codons  and anticodons  in $5' \to 3'$ direction. As an anticodon is antiparallel to codon, the 1st nucleotide (respectively the 3rd nucleotide) of the anticodon is paired to the 3rd (respectively the 1st) nucleotide of the codon.  

The Watson-Crick codon-anticodon recognition pattern is not the most economical one, as it needs 61 anticodons and
Nature, through the wobble mechanism, makes us of a lower number of anticodons and we will discuss the real pattern in the next Section. 

In our approach the  codon-anticodon interaction 
 is given by the ``spin-spin" interaction 
 \begin{center}
 $T_{X^cZ^cN^c,N'^aZ^a_cX^a_c} \equiv T_{NN'} $ 
 \end{center}
 introduced in \cite{SS12}, see Appendix 2 for a short reminder.

So  we write $\wh{P}_N$ as
\be
 \wh{P}_N \equiv F(T_{NN_c}) \equiv F(K_N)
\ee 
where
\be
K_N \equiv n_H \, c_H + n_V \, c_V
\ee
where $n_H, n_V$ are positive or negative integers depending on the codon-anticodon pattern, determined by the interaction and computed in Tables 3 and 4 of \cite{SS12} and  $c_H$ and $c_V$  are two real, theoretically indeterminable, parameters of the interaction. As ``smooth" behavior of $F(K_N)$, we assume that the signs and, possibly, some inequalities between the values of the parameters  $c_H$ and $c_V$ are independent from the biological species (at least inside a suitable domain of biological species with almost equal GC content) and 
dependent on the di-nucleotide $XZ$. Actually we do not need the knowledge of the values of
  $c_H$ and $c_V$, but their sign and some inequalities between their values.
  
 In the previous papers we have made the assumption that  $c_H$ and $c_V$ were depending on the encoded amino acid, but we were essentially considering quartets. 

Let us repeat, once again, that, in order not to make the notation heavier, we do not explicitly write the dependence on the di-nucleotide, when not necessary.

As we have assumed that $F(K_N)$ is a decreasing function in the interaction, determined by several conditions (minimization of energy, maximization of entropy, abundance or preference of specific anticodon, etc.),
 then it follows that the largest value 
of $\wh{P}_{XZN}$ corresponds to the $XZN$ with the smallest value of $F(K_N)$, that is with the smallest value $K_N$ and so on. 

In Table \ref{Table:Inter-WC} we explicitly report the value of the coefficients of this interaction computed from our model in Table 3  and Table 4 of \cite{SS12}.\footnote{In Table 4 of \cite{SS12} due to a misprint the doublet AGR is missing.}

\begin{table}[H]
% \footnotesize
 \begin{tabular}{|c||c|c|c|c|}
 \hline
codon & GCU & GCA & GCC & GCG \\
  \hline
 $T$ & ${\bf -10c_H - 22c_V} $& -10$c_H$ + 2$c_V $ & 18$c_H$ - 22$c_V$ & 18$c_H$ + 2$c_V$ \\ \hline
        codon & CGU & CGA & CGC & CGG \\ \hline
        $T$ & -10$c_H$ - 6$c_V$ & ${\bf -10c_H + 2c_V}$ & 18$c_H$ - 6$c_V$ & 18$c_H$ + 2$c_V$ \\ \hline
        codon & CCU & CCA & CCC & CCG \\ \hline
        $T$ & ${\bf -10c_H - 30c_V} $& -10$c_H$ - 6$c_V $ & 18$c_H$ - 30$c_V$ & 18$c_H$ - 6$c_V$  \\ \hline
        codon & GGU & GGA & GGC & GGG  \\ \hline
        $T$ & -10$c_H$ - 6$c_V$ & ${\bf -10c_H +18c_V}$ & 18$c_H$ - 6$c_V$ & 18$c_H$ + 18$c_V$ \\ \hline
        codon & CUU & CUA & CUC & CUG \\  \hline
        $T$ & ${\bf 6c_H - 30c_V} $& 6$c_H$ - 6$c_V $ & 2$c_H$ - 30$c_V$ & 2$c_H$ - 6$c_V$ \\ \hline
        codon & GAU & GAA & GAC & GAG \\ \hline
        $T$ & 6$c_H$ - 6$c_V$ & ${\bf 6c_H + 18c_V} $ & 2$c_H$ - 6$c_V$ & 2$c_H$ + 18$c_V$ \\ \hline
        codon & GUU & GUA & GUC & GUG  \\ \hline
        $T$ & ${\bf 6c_H - 22c_V} $& 6$c_H$ + 2$c_V $ & 2$c_H$ - 22$c_V$ & 2$c_H$  + 2$c_V$ \\ \hline
        codon & CAU & CAA & CAC & CAG \\ \hline
        $T$ & 6$c_H$ - 6$c_V$ & ${\bf  6c_H + 2c_V}$  & 2$c_H$ - 6$c_V$ & 2$c_H$ + 2$c_V$  \\ \hline
        codon & UCU & UCA & UCC & UCG  \\ \hline
        $T$ &$ {\bf 2c_H - 30c_V}$ & 2$c_H$ - 6$c_V $ & 6$c_H$ - 30$c_V$ & 6$c_H$ - 6$c_V$ \\ \hline
        codon & AGU & AGA & AGC & AGG \\  \hline
        $T$ & 2$c_H$ - 6$c_V$ & ${\bf 2c_H + 18c_V}$  & 6$c_H$ - 6$c_V$ & 6$c_H$ + 18$c_V$  \\ \hline
        codon & UUU & UUA & UUC & UUG \\ \hline
        $T$ & ${\bf 18c_H - 30c_V}$ & 18$c_H$ - 6$c_V $ & -10$c_H$ - 30$c_V$ & -10$c_H$ - 6$c_V$ \\ \hline
        codon & AAU & AAA & AAC & AAG \\  \hline
        $T$ & 18$c_H$ - 6$c_V$ & ${\bf 18c_H + 18c_V}$  & - 10$c_H$ - 6$c_V$ & -10$c_H$ + 18$c_V$ \\ \hline
        codon & ACU & ACA & ACC & ACG \\ \hline
        $T$ & ${\bf  2c_H - 22c_V}$ & 2$c_H$ + 2$c_V$ & 6$c_H$ - 22$c_V$ & 6$c_H$ + 2$c_V$ \\ \hline
        codon & AUU & AUA & AUC & AUG \\ \hline
        $T$ & ${\bf 18c_H - 22c_V}$ & 18$c_H$ + 2$c_V $ & -10$c_H$ - 22$c_V$ & -10$c_H$ + 2$c_V$ \\ \hline
 \end{tabular} \centering\caption{Value of the  codon-anticodon interaction $T$ according to the Watson-Crick recognition pattern. In bold the smallest value of the interaction with the signs of $c_H$ and $c_V$ reported below in Table \ref{Table:rank-WC}. }
 \label{Table:Inter-WC}
 \end{table}

From the previous Table we see that the smallest value of the interaction is determined only from the signs of the parameters  $c_H$ and $c_V$.

Recalling that $J_{H,3}$ ($J_{V,3}$)  of the nucleotide $X$ is equal (respectively opposite) to $J_{H,3}$ ($J_{V,3}$)  of  the  nucleotide $X_c$, see Appendix 2, from the property of the tensor product  \cite{K} of the conjugate triplets of nucleotides it follows immediately that the numerical coefficients of $c_H$ are equal for the two conjugate quadruples (as an example CGN conjugate to GCN).

\begin{table}[H]
% \scriptsize
 \centering
 \begin{tabular}{|c||c|c|c|c|}
 \hline
Quadruples & Hierarchy & sign $c_H$ & sign $c_V$ & Constraint \\ \hline
GCN &  $P_U > P_A > P_C > P_G$ (19) & + & + & $7 c_H > 6 c_V$ \\
 \hline 
CGN &  $P_A > P_U > P_G > P_C$(11)& + & - &$ 7 c_H > 2 |c_V|$ \\ 
\hline \hline
CCN  & $P_U > P_A > P_C  > P_G$ (19) & + & +  & $7 c_H > 6 c_V$ \\ 
\hline
GGN & $P_A > P_U > P_G  > P_C$  (20) & + & -  & $7 c_H > 6 |c_V|$  \\
  \hline \hline
  CUN & $P_U  > P_A > P_C  > P_G$(17) & - & + & $|c_H|  >  6 c_V$ \\ 
  \hline
GAN & $P_A > P_U > P_G  > P_C$ (20) & - & - & $ |c_H| >  6|c_V|$ \\
  \hline \hline
 GUN & $P_U > P_A > P_C > P_G$  (1) & - & + & $|c_H| > 6 c_V$ \\
\hline
CAN & $P_A > P_U > P_G  > P_C$ (20) & - & - & $|c_H|  > 2|c_V|$ \\
 \hline \hline
UCN & $P_U > P_A > P_C  > P_G$ (17) & + & + & $c_H > 6 c_V$ \\
\hline
AGN & $P_A > P_U > P_G  > P_C$ (19) & + & - & $c_H > 6 |c_V|$ \\
 \hline \hline
UUN & $P_U > P_A > P_C  > P_G$ (2) & - & + & $7|c_H| > 6 c_V$ \\
\hline
AAN & $P_A > P_U > P_G  > P_C$ (17) & - & - & $7|c_H| >  6|c_V|$ \\ 
\hline \hline
ACN & $P_U > P_A > P_C  > P_G$ (20)& + & + & $c_H > 6 c_V$ \\
 \hline \hline
AUN &  $P_U > P_A > P_C  > P_G$  (1) & - & + & $7|c_H| > 6 |c_V|$ \\
 \hline
\end{tabular}\centering\caption{Hierarchy of the codon usage frequencies, signs and constraints on the constants  $c_H$ and $c_V$. In brackets, the number of cases where the observed hierarchy in the presently considered specimen is equal to the theoretical computed one. }
 \label{Table:rank-WC}
 \end{table}
 
 Note that the quadruplets UGN and UAN  
conjugate to ACN and AUN respectively, are not considered since they
include Stop codons.
 For any quadruple, the hierarchy of  the values of the interaction, which determines the hierarchy of $\wh{P}_N$ and, consequently, from {\bf Assumption 2} of Sec. 2 that of $P_N$, is computed from the signs of $c_H$ and $c_V$ and from the inequalities between their values, reported, respectively, in the second and in the last column of Table \ref{Table:rank-WC}.

We remark, from Table \ref{Table:rank-WC},  that the parameters  $c_H$ and $c_V$  of two conjugate quadruples 
 have, respectively, the same and the opposite sign and that the corresponding hierarchy of $P_N$ is the conjugate one, i.e
 $P_{XZU} > P_{XZA}  > P_{XZC} > P_{XZG}$  and $P_{X_cZ_cA} > P_{X_cZ_cU}  > P_{X_cZ_cG} > P_{X_cZ_cC}$. 
 
\medskip

Let us note that:
\begin{itemize}
\item changing the sign of $c_V$ in Table \ref{Table:rank-WC} (and, if it is the case, $c_V \ra |c_V|$ in the constraint), for all the quadruples, the hierarchy goes into its conjugate one, i.e. ($P_U > P_A > P_C  > P_G) \; \llra  \; (P_A > P_U > P_G  > P_C$).

\item some hierarchies, in particular $P_U > P_A > P_G  > P_C$ and   $P_A > P_U > P_C  > P_G$, are not consistent with the form of the interaction. 
 \end{itemize}

 We remark that, in the Watson-Crick codon-anticodon pattern, we get the theoretical hierarchies of $P_N$, which, in most cases, correspond to the observed one, from the following symmetry:
 
 \medskip
 
 {\noindent \bf  The parameters $c_H$ are positive (negative) for the di-nucleotides appearing in the left half  (respectively right) of the matrix $\mathbb{B}_1 \otimes \mathbb{B}_2$, see eq.(\ref{eq:array1}) of Appendix, and the parameters $c_V$ are positive  (negative) for the di-nucleotides appearing in the upper half (respectively lower) of the matrix. It follows that the constants  $c_H$ and $c_V$  of two conjugate quadruples have respectively the same and the opposite sign.} 
 
 \section{The model for the real codon-anticodon recognition pattern}

The  codon-anticodon recognition pattern, which reveals in the largest majority of the analyzed species,  is summarized in Table \ref{Table:pat-c-a}, from the data reported in literature \cite{Salvia}-\cite{Glyc}.  A small number of changes from the pattern reported in Table \ref{Table:pat-c-a} appears in the literature and we will separately discuss them  below.

\begin{table}[H]
 \footnotesize
 \begin{tabular}{|c|c||c|c|c|}
 \hline
Codon & Anticodon &  Codon & Anticodon \\
\hline
CGN & ACG & GCN & UGC \\
\hline
\hline
CCN & UGG  & GGY & GCC \\
 &.           & GGR & UCC \\
\hline
\hline
CUN & UAG & GAY & GUC \\
 &. & GAR & UUC \\
\hline
\hline
 GUY  & GAC &  CAY & GUG \\
 GUR & UAC & CAR & UUG \\
 \hline\hline
 UCY  & GGA & AGY &  GCU \\
 UCR  & UGA & AGR &  UCU \\
\hline
\hline
 UUY  & GAA & AAY & GUU \\
 UUG & CAA   & AAR & UUU \\
 UUA & UAA & & \\
\hline
\hline
 ACY & GGU &  AUY & GAU \\
 ACR & UGU & AUR & CAU \\
\hline
\hline
\end{tabular}\centering\caption{Pattern of codon-anticodon interaction inferred  from the cited literature.}
 \label{Table:pat-c-a}
 \end{table}

In Table \ref{Table:Inter} we report the value of the interaction $T$ for any codon, computed from Table 3  and Table 4 of \cite{SS12}, according the the codon-anticodon pattern of Table \ref{Table:pat-c-a}.

\begin{table}[H]
 \footnotesize
 \begin{tabular}{|c||c|c|c|c|}
 \hline
codon & GCU & GCA & GCC & GCG \\ \hline
        $T$ & -$10c_H + 6c_V$ & -10$c_H$ + 2$c_V $ & -6$c_H$ + 6$c_V$ & -6$c_H$ + 2$c_V$ \\ \hline
        codon & CGU & CGA & CGC & CGG \\ \hline
        $T$ & -10$c_H$ - 6$c_V$ &  $-10c_H + 2c_V$ & -6$c_H$ - 6$c_V$ & -6$c_H$ + 2$c_V$ \\ \hline
        codon & CCU & CCA & CCC & CCG \\ \hline
        $T$ & $ -10c_H - 10c_V$ & -10$c_H$ - 6$c_V $ & -6$c_H$ - 10$c_V$ & -6$c_H$ - 6$c_V$ \\ \hline
        codon & GGU & GGA & GGC & GGG \\ \hline
        $T$ & 6$c_H$ - 6$c_V$ & $ -10c_H +18c_V$ & 18$c_H$ - 6$c_V$ & -6$c_H$ + 18$c_V$ \\ \hline
        codon & CUU & CUA & CUC & CUG \\ \hline
        $T$ & 6$c_H$ - 10$c_V$ & 6$c_H$ - 6$c_V $ & -10$c_H$ - 10$c_V$ & -10$c_H$ - 6$c_V$ \\ \hline
        codon & GAU & GAA & GAC & GAG \\ \hline
        $T$ & 2$c_H$ - 6$c_V$ & 6$c_H$ + 18$c_V $ & 2$c_H$ - 6$c_V$ & -10$c_H$ + 18$c_V$ \\ \hline
        codon & GUU & GUA & GUC & GUG \\ \hline
        $T$ & 2$c_H$ - 22$c_V$ & 6$c_H$ + 2$c_V $ & 2$c_H$ - 22$c_V$ & -10$c_H$  + 2$c_V$ \\ \hline
        codon & CAU & CAA & CAC & CAG \\ \hline
        $T$ & 2$c_H$ - 6$c_V$ & 6$c_H$ + 2$c_V $ & 2$c_H$ - 6$c_V$ & -10$c_H$ + 2$c_V$ \\ \hline
        codon & UCU & UCA & UCC & UCG \\ \hline
        $T$ & 2$c_H$ - 30$c_V$ & 2$c_H$ - 6$c_V $ & 6$c_H$ - 30$c_V$ & -10$c_H$ - 6$c_V$ \\ \hline
        codon & AGU & AGA & AGC & AGG \\ \hline
        $T$ & 2$c_H$ - 6$c_V$ & 2$c_H$ + 18$c_V $ & 6$c_H$ - 6$c_V$ & -10$c_H$ + 18$c_V$ \\ \hline
        codon & UUU & UUA & UUC & UUG \\ \hline
        $T$ & 6$c_H$ - 30$c_V$ & 18$c_H$ - 6$c_V $ & -10$c_H$ - 30$c_V$ & -10$c_H$ - 6$c_V$ \\ \hline
        codon & AAU & AAA & AAC & AAG \\ \hline
        $T$ & 6$c_H$ - 6$c_V$ & 18$c_H$ + 18$c_V $ & - 10$c_H$ - 6$c_V$ & -6$c_H$ + 18$c_V$ \\ \hline
        codon & ACU & ACA & ACC & ACG \\ \hline
        $T$ & 2$c_H$ - 22$c_V$ & 2$c_H$ + 2$c_V $ & 6$c_H$ - 22$c_V$ & -10$c_H$ + 2$c_V$ \\ \hline
        codon & AUU & AUA & AUC & AUG \\ \hline
        $T$ & 6$c_H$ - 22$c_V$ & 6$c_H$ + 2$c_V $ & -10$c_H$ - 22$c_V$ & -10$c_H$ + 2$c_V$ \\ \hline
 \end{tabular} \centering\caption{Value of the interaction codon-anticodon according to the pattern specified in Table \ref{Table:pat-c-a}.}
 \label{Table:Inter}
 \end{table}
 
The quartet Gly (GGN) for the biological species {\it Ageratina adenophora}, {\it Oryza minuta}, {\it Amonum compactum},  {\it Morus cathayana}  and {\it Glycine soja} is read only by the anticodon UCC, while for {\it Ulva flexuosa} it is read by the anticodon GCC. Therefore for these biological species the line relative to Gly in Table {\ref{Table:Inter}} has to be replaced by the Table \ref{Table:Gly-1}

\begin{table}[H]
 \footnotesize
 \begin{tabular}{|c||c|c|c|c||c|}
 \hline
 codon &  GGU & GGA & GGC & GGG  & anticodon \\
 \hline
 $T$ &-10$c_H$ + 18$c_V$ & -10$c_H$ + 18$c_V $ & -6$c_H$ + 18$c_V$ &  -6$c_H$ + 18$c_V$  & UCC \\
 \hline
 \hline
 codon &  GGU & GGA & GGC & GGG & anticodon \\
 \hline
 $T$ & 6$c_H$ - 6$c_V$  &  6$c_H$ + 18$c_V $ &18$c_H$ - 6$c_V$ &   18$c_H$ + 18$c_V$ & GCC  \\
\hline
 \end{tabular} \centering\caption{Value of the interaction codon-anticodon for {\it Ageratina adenophora}, {\it Oryza minuta}, {\it Amonum compactum},  {\it Morus cathayana} and {\it Glycine soja} (upper part), for {\it Ulva flexuosa} (lower part).}
 \label{Table:Gly-1}
 \end{table}
 
 In the above Tables the smallest value of the interaction is not generally determined only by the signs of $c_H$ and $c_V$, but in some case there is degeneracy, see discussion below, and in other cases constraints are required between the values of $c_H$ and $c_V$.

For  {\it Cocos nucifera} the codon UUA is also read by the anticodon CAA, therefore the entry 18$c_H$ - 6$c_V$  in Table {\ref{Table:Inter}} relative to this codon should be replaced by 6$c_H$ - 6$c_V$.  

Looking at Table \ref{Table:Inter} and Table \ref{Table:Gly-1}, we are faced with cases  where  the value of $K_N$ is the same for two codons in the same quadruple, i.e. in Table  \ref{Table:Inter} GAC-GAU, GUC-GUU and CAC-CAU, in Table   \ref{Table:Gly-1} GGC-GGG and GGU-GGA. In all these cases the data show that the value of usage frequency of the codon ending with U is higher than the one of the codon the ending with C, $P_U > P_C$, and, for Gly, $P_A > P_U, P_G > P_C$. According to the idea underlying our approach we look for an explication rooted in the structure of codon-anticodon interaction.

So we add a further (symmetric) term to the codon-anticodon interaction, which has been already suggested in \cite{SS12},  and which we will call the z-spin interaction between the codon $X^cZ^cN^c$ and the anticodon $N'^aZ^a_cX^a_c$
\be
T^z_{X^cZ^cN^c,N'^aZ^a_cX^a_c} \equiv T^z_{N^cN'^a} =   4 \, g_H \, J^{c}_{H,3} \, J^{a}_{H,3} \; + \;  4 \, g_V \, J^{c}_{V,3} \, J^{a}_{V,3}
\label{eq:zz}
\ee
where $g_H$ and $g_V$ are real indeterminate parameter.
 This term has not been introduced in the previous papers and we have above explained the reasons to do it now. 

So the total interaction between the codon and the anticodon now reads
\be
T'_{NN'} =  T_{NN'} + T^z_{NN'}
\label{eq:T-j}
\ee
where the expression of $T_{NN'}$ is reported in the Appendix 2.
Using the additivity of $J_{H,3}$ and $J_{V,3}$ and the property recalled in Section 3 of  the Watson-Crick conjugate nucleotide $X_c$ we can rewrite the expressions which appear in eq.(\ref{eq:zz}) in the following way:
\bea
J^{c}_{H,3} \, J^{a}_{H,3}  & =  &  (J^{XZ}_{H,3})^2 
+ J^{XZ}_{H,3} (J^N_{H,3} + J^{N\prime}_{H,3}) + J^N_{H,3} J^{N\prime}_{H,3} \label{eq:zzH} \\
J^{c}_{V,3} \, J^{a}_{V,3}  & =  & - (J^{XZ}_{V,3})^2 
+ J^{XZ}_{V,3} (J^{N\prime}_{H,3} - J^N_{H,3}) + J^N_{V,3} J^{N\prime}_{V,3} 
\label{eq:zzV}
\eea
where we denote with $J^{XZ}_{\alpha,3}$ ($J^N_{\alpha,3}$) the third component of $\vec{J}_{\alpha}$ ($\alpha$ = H, V) of the di-nucleotide  XZ (respectively of nucleotide N).
The first term of eqs.(\ref{eq:zzH})-(\ref{eq:zzV}), being equal for all the codons in the quadruples, will be omitted in the following as it is irrelevant for our considerations.

  So we write
\be
\wh{P}_N \equiv F( T_{NN'} + T^z_{NN'}) \equiv F(K_N + m_H g_H +  m_V g_V)
\ee
where $m_H, m_V$ are positive or negative integers, determined by the the crystal basis model, and computed in Tables \ref{Table:JJ} and \ref{Table:JJ-1}. 

\begin{table}[H]
 \footnotesize
 \begin{tabular}{|c||c|c|c|c|}
 \hline
        codon & GCU & GCA & GCC & GCG \\ \hline
        $J$ & -3$g_H$ + $g_V$ & ${\bf -3g_H - g_V}$ & -$g_H$ + $g_V$ & -$g_H$ - $g_V$ \\ \hline
        codon & CGU & CGA & CGC & CGG \\ \hline
        $J$ &  ${\bf -3g_H - g_V}$ & -3$g_H$ + $g_V $ & -$g_H$ - $g_V$ & -$g_H$ + $g_V$ \\ \hline
        codon & CCU & CCA & CCC & CCG \\ \hline
        $J$ &  ${\bf -3g_H + g_V}$ & -3$g_H$ + 3$g_V $ & -$g_H$ + $g_V$ & -$g_H$ + 3$g_V$ \\ \hline
        codon & GGU & GGA & GGC & GGG \\ \hline
        $J$ & -$g_H$ + 3$g_V$ &  ${\bf -3g_H - 5g_V}$ & 5$g_H$ + 3$g_V$ & -$g_H$ - 5$g_V$ \\ \hline
        codon & CUU & CUA & CUC & CUG \\ \hline
        $J$ & $g_H$ + $g_V$ & $g_H$ + 3$g_V $ &  ${\bf -g_H + g_V}$ & -$g_H$ + 3$g_V$ \\ \hline
        codon & GAU & GAA & GAC & GAG \\ \hline
        $J$ & -$g_H$ + 3$g_V$ & $g_H$ - 5$g_V $ & $g_H$  + 3$g_V$ & ${\bf -g_H - 5g_V}$ \\ \hline
        codon & GUU & GUA & GUC & GUG \\ \hline
        $J$ &  ${\bf -g_H - g_V}$ & $g_H$ - $g_V $ & $g_H$ - $g_V$ &  ${\bf -g_H  - g_V}$ \\ \hline
        codon & CAU & CAA & CAC & CAG \\ \hline
        $J$ & ${\bf -g_H - g_V}$ & $g_H$ - $g_V $ & $g_H$ - $g_V$ & ${\bf -g_H - g_V}$ \\ \hline
        codon & UCU & UCA & UCC & UCG \\ \hline
        $J$ & ${\bf -g_H - 5g_V}$ & $g_H$ + 3$g_V $ & $g_H$ - 5$g_V$ & -$g_H$ + 3$g_V$ \\ \hline
        codon & AGU & AGA & AGC & AGG \\ \hline
        $J$ & -$g_H$ + 3$g_V$ & $g_H$ - 5$g_V $ & $g_H$ + 3$g_V$ & ${\bf -g_H - 5g_V}$ \\ \hline
        codon & UUU & UUA & UUC & UUG \\ \hline
        $T$ & ${\bf -g_H - 5g_V}$ & 5$g_H$ + 3$g_V $ & -3$g_H$ - 5$g_V$ & -3$g_H$ + 3$g_V$ \\ \hline
        codon & AAU & AAA & AAC & AAG \\ \hline
        $J$ & -$g_H$ + 3$g_V$ & 5$g_H$ - 5$g_V $ & - 3$g_H$ + 3$g_V$ & ${\bf -g_H - 5g_V}$ \\ \hline
        codon & ACU & ACA & ACC & ACG \\ \hline
        $J$ & ${\bf -g_H - g_V}$ & $g_H$ - $g_V $ & $g_H$ - $g_V$ & ${\bf -g_H - g_V} $ \\ \hline
        codon & AUU & AUA & AUC & AUG \\ \hline
         $J$ & -$g_H$ - $g_V$ & -$g_H$ - $g_V $ &   ${\bf -3g_H - g_V}$ &  ${\bf -3g_H - g_V}$ \\ \hline
 \end{tabular} \centering\caption{Value of the relevant z-spin interaction between codon-anticodon according to the pattern specified in Table \ref{Table:pat-c-a}. In bold the smallest value of the interaction with positive signs of $g_H$ and $g_V$.}
 \label{Table:JJ}
 \end{table}

\begin{table}[H]
 \footnotesize
 \begin{tabular}{|c||c|c|c|c|}
\hline
codon &  GGU & GGA & GGC & GGG  \\
\hline
$J$ & -3$g_H$ + $g_V$ &  ${\bf -3g_H - 5g_V}$  & -$g_H$ + $g_V$ & -$g_H$ - 5$g_V $ \\
\hline
\hline
codon &  GGU & GGA & GGC & GGG  \\
\hline
$J$ & -$g_H$ + 3$g_V$ & ${\bf -g_H + g_V}$  & 5$g_H$ + 3$g_V$ & 5$g_H$ +$g_V $ \\
\hline
 \end{tabular} \centering\caption{Value of the relevant z-spin interaction for Gly according to 
 codon-anticodon pattern of Table 2.}
 \label{Table:JJ-1}
 \end{table}

For  {\it Cocos nucifera} the coefficient 5$g_H$ + 3$g_V $ in Table \ref{Table:JJ} should be replaced by -$g_H$ + 3$g_V$.
 
 We remark that  if $g_H > 0$ and $g_V > 0$
  the degeneracy is removed in the good direction for GAC-GAU, GUC-GUU, CAC-CAU, GGC-GGG and GGU-GGA.
  So we assume that, for all the quadruples,  $g_H > 0$ and  $g_V > 0$. We remark that in the Watson-Crick recognition pattern there is no degeneracy in the values of the interaction, so we do not need to discuss the z-spin interaction for this pattern. However, adding this term to the values of Table \ref{Table:Inter-WC} does not change in a significant way the results of the previous Section.
 
\section{Discussion of the model}

\bigskip

 In the following, we analyze, in detail, the couples of conjugates quadruples and we summarize the obtained results in Table \ref{Table:sign}.
 
Also in this case,  the hierarchy of  the values of the interaction, determined by $K_N + m_H g_H +  m_V g_V$, is computed from the signs of $c_H$ and $c_V$, from the inequalities between their values, see Table \ref{Table:sign} and the following paragraph, and from the positive signs of $g_H$ and $g_V$.  The hierarchy of these values determines the hierarchy of $\wh{P}_N$ and, consequently, from {\bf Assumption 2} of Sec. 2 that of $P_N$.

  Let us emphasize, once more, that the rank of values of the interaction, once fixed $c_H$ and $c_V$  and $g_H$ and
  $g_V$, follows from the values of the numerical coefficients which derive from the model. 
 
\begin{table}[H]
 %\scriptsize
 \begin{tabular}{|c||c|c|c|c|}
 \hline
Quadruples & Hierarchy & sign $c_H$ & sign $c_V$ & Constraints \\ \hline
GCN &  $P_U > P_A > P_C > P_G$ (19) & + & - &$2 c_H + g_H > 2 |c_V| - g_V > 0$ \\ \hline 
CGN &  $P_A > P_U > P_G > P_C$ (11) & + & - &$2 c_H + g_H > 4 |c_V| - g_V > 0$ \\ 
& $P_U > P_A > P_C  > P_G$ (1) & + & - & $2 c_H + g_H  >  g_V - 4 |c_V| > 0$ \\ 
 \hline \hline
CCN  & $P_U > P_A > P_C  > P_G$ (19) & + & +  & $2 c_H  + g _H > 2 c_V + g_V$ \\ \hline
GGN & $P_A > P_U > P_G  > P_C$ (20)  & + & +  & $8c_H + g_H > 12 c_V - 4g_V >  6c_H$ \\
  \hline \hline
  CUN & $P_U  > P_A  > P_C  > P_G$ (17) & - & - & $8|c_H| - g_H >  g_V - 2 |c_V|$ \\ 
    & & & & $g_V > 2|c_V|$  \\ \hline
GAN & $P_A > P_U > P_G  > P_C$ (20) & - & - & $6 |c_H| >  12 |c_V|  + 4 g_V > 6 |c_H| - g_H$ \\
& & & & $4 |c_H| > g_H$\\  \hline \hline
 GUN & $P_A > P_U > P_G  > P_C$ (15) & - & - & $2|c_V| > |c_H|,\; 8|c_H| > g_H$ \\ 
 & $P_U > P_A > P_C  > P_G$ (1) & - & - & $2 |c_H| - g_H  > 12 |c_V|$ \\ \hline
 CAN  & $P_A > P_U > P_G  > P_C$ (20) & - & - & $6|c_H| >  4|c_V| > 6|c_H| -  g_H > $  \\
& & & & $4|c_V| >  g_H - 2|c_H|$ \\ \hline \hline
UCN & $P_U > P_A > P_C  > P_G$ (15) & - & - & $8|c_H| - g_H > 12 |c_V|  - 4 g_V > 2|c_H|$ \\ 
& & & & $g_H > 12 |c_V|  - 4 g_V$ \\ \hline
AGN & $P_A > P_U > P_G  > P_C$ (19) & - & - & $6|c_H| > 12 |c_V|  + 4 g_V > 8|c_H| - g_H$ \\
& & & &$12 |c_V|  + 4 g_V  > g_H$  \\ \hline \hline
UUN & $P_U > P_A > P_G  > P_C$ (17) & - & - & $3 g_H -6|c_H|  > 12 |c_V| - 
4 g_V$ \\
& & & & $3|c_V| > g_V$, $7|c_H| >  2 g_H$\\ \hline
AAN & $P_A > P_U > P_G  > P_C$ (17) & - & - & $3|c_H| >  6|c_V| +  2g_V >  3/2 g_H - 3 |c_H|$ \\
 \hline \hline
ACN & $P_U > P_A > P_C  > P_G$ (20) & - & - &  $8|c_H| - g_H >  12|c_V| >  2|c_H|$ \\
& & & &$g_H > 12|c_V|$\\ \hline \hline
AUN &  $P_U > P_A > P_C  > P_G$ (1) & - & + & $8|c_H|  - g_H >  12 c_V$ \\ 
\hline
\end{tabular}
\centering
\caption {For each quadruple, the resulting hierarchy of the codon usage probabilities computed assuming that the constants $c_H$ and $c_V$ have the sign and satisfy the constraints specified in the table. The number in brackets specifies how many times the hierarchy is observed in the experimental data reported in the Tables \ref{Table:Sal-Coc}-\ref{Table:Epip-Glyc}.}
\label{Table:sign}
\end{table}

For Gly (GGN), for the five biological species listed in the Table \ref{Table:Gly-1}, and  for UUN for {\it Cocos nucifera} we get, respectively, the constraints $4 c_H + 2 g_H > 6 g_V$ and $g_V > 3 |c_V|$,  $8 |c_H| - g_H > 4 g_V - 12 |c_V|$.

Let us summarize our main results:
 
 \begin{itemize}
 
 \item the  constants $c_H$  and $c_V$ have the same sign for two conjugate quadruples.
 
 \item the quadruples  GCN and CGN  have the unique peculiarity that the codons are read, for each quadruple, only by one anticodon (respectively ACG and UGC) which are conjugate each other.
 
 \item for UUN the hierarchy $P_U > P_A > P_G  > P_C$ is compatible with our model due to the z-spin term.
 
 \item for CGN, GUN, UUN and AUN,  the  hierarchy $P_U > P_A > P_G  > P_C$  (observed  respectively in 8, 6,17 and 19 in biological species)  is not compatible with the spin-spin and z-spin interaction. Let us recall that the same behavior was present in the Watson-Crick recognition pattern.
 
\end{itemize}

In Table \ref{Table:l-pl} we list the considered 20 plants and in Tables  \ref{Table:st-h} and  \ref{Table:ex-h}  we report, for any plant, the observed hierarchy of the codon usage frequencies. 

% \newpage
 
  \begin{table}[H]
   %\footnotesize
 \begin{tabular}{|c||c||c||c|}
\hline
N.& Biological species & N.  &  Biological species \\
\hline
1 & {\it Salvia mitiorrhiza} & 11 &{\it Oryza minuta}  \\
2 & {\it Cocos nucifera} & 12  & {\it Citrus aurantiifolia} \\
3 & {\it  Artemisia selengensis} &  13 & {\it Amonum compactum} \\
4 & {\it Utricularia reniformis} & 14 &  {\it Morus cathayana} \\
5 & {\it  Cynara cardunculus} &15 & {\it Castanopsis sclerophylla} \\
6 & {\it Aconitum pseudolaeve} & 16 &   {\it Zingiber montanum}\\
7 & {\it Arabidopsis thaliana} & 17 & {\it Musa acuminata} \\
8 & {\it Ageratina adenophora} & 18 & {\it Brassica oleracea} \\
9 & {\it Lotus japonicus} & 19 & {\it Epipremnum aureum} \\
10 & {\it Phoenix dactylifera L.} & 20 & {\it Glycine soja} \\
\hline
 \end{tabular} \centering\caption{List of the considered biological species.}
 \label{Table:l-pl}
 \end{table}
 
 \begin{table}[H]
 \footnotesize
 \begin{tabular}{|c||c|c|c|c|}
\hline
Quadruples & $P_U > P_A > P_C > P_G$ &  $P_A > P_U > P_G > P_C$ & $P_U > P_A > P_G > P_C$ &  $P_A > P_U > P_C > P_G$\\
\hline
GCN & 1-20& $\sim$ &  $\sim$&  $\sim$\\
\hline
CGN & $\sim$   & 1-4, 9-12, 14-16, 18, 20 &5, 6, 8, 13, 17, 19  & 7 \\
\hline
CCN &1, 2, 4-20 & $\sim$ & $\sim$ &3\\
\hline
GGN &$\sim$  & 1-20 &$\sim$  &$\sim$  \\
\hline
CUN &2, 5-20 & $\sim$  &1, 4 & $\sim$ \\
\hline
GAN & $\sim$ &1-20 & $\sim$ & $\sim$ \\
\hline
GUN &3  & 2, 4-6, 8, 11, 12, 14-17, 20  & 1, 7, 9, 10, 13, 18 &  19 \\
\hline
CAN & $\sim$  & 1-20 & $\sim$ & $\sim$ \\
\hline
UCN & 1, 2, 4-10, 12-14, 17, 19, 20  & $\sim$ & 15, 16 &  18 \\
\hline
AGN & $\sim$  & 1-20  & $\sim$ & $\sim$ \\
\hline
UUN & 2, 13  & $\sim$ & 1, 4-12, 14-20 &  $\sim$ \\
\hline
AAN &  $\sim$ & 1, 2, 4-18  & 3, 19, 20 & $\sim$ \\
\hline
ACN & 1, 2, 4-20 & $\sim$ &  $\sim$ &  $\sim$\\
\hline
AUN & 3 & $\sim$ & 1, 2, 4-10, 12-20 & $\sim$ \\
\hline
 \end{tabular} \centering\caption{The most observed  hierarchies of the codon usage frequencies for fourteen quadruples, the  number identifies the biological species according to Table \ref{Table:l-pl}.}
 \label{Table:st-h}
 \end{table}

\begin{table}[H]
 \footnotesize
 \begin{tabular}{|c||c|c|c|}
\hline
Quadruples & $P_U > P_C > P_G > P_A$ & $P_U > P_C> P_A > P_G$  &  $P_U > P_G > P_A > P_C$ \\
\hline
CUN & 3 & $\sim$ &  $\sim$\\
\hline
UCN &3 & 11 &  $\sim$\ \\
\hline
UUN & 3 & $\sim$ & $\sim$ \\
\hline
ACN &$\sim$  & 3 &$\sim$   \\
\hline
AUN &$\sim$   & $\sim$   & 11 \\
\hline
 \end{tabular} \centering\caption{The observed exceptional hierarchies  of the  codon usage frequencies for fourteen quadruples, the  number identifies the biological species according to Table \ref{Table:l-pl}.}
 \label{Table:ex-h}
 \end{table}

 In the majority of cases, the disagreement between the predicted and the observed hierarchy appears with the pattern  $P_U > P_A > P_G  > P_C$, in particular for  AUN. This hierarchy is not compatible with the spin-spin and z-spin interaction as it would require $c_V > 0$ and in the same time $c_V < 0$. Let us recall that the same behavior was present with the Watson-Crick recognition pattern. However, let us note that the disagreement between the predicted and the observed hierarchy is on the last two codon usage frequencies (predicted  $P_C  > P_G$, observed  $P_G  > P_C$), where we can expect that, being lower the value of $\wh{P}_N$, the term of $\rho_N$ may modify the order.

The constraints on the interaction parameters in the real codon-anticodon recognition pattern are more complicated than in the case of the Watson-Crick pairing and in some case require a fine tuning, which was unnecessary  in the Watson-Crick recognition pattern.

We must emphasize the oversimplifications of our model:
\begin{itemize}

\item The nucleotides which, through the ``wobble mechanism”,  recognize and bind to more than its conjugate ones present
chemical modifications which we do not take into account and assign the same  ``labels" than the unmodified one, consequently the same interaction.

\item Same anticodons appear in different tRNA genes which presumably behave in different way. We do not consider this aspect as the codon-anticodon interaction in the crystal basis model is between the triples of nucleotides, i.e., in some sense, a  ``local" one.
\end{itemize}

However we have to consider that, besides the remarks above reported,  in the real codon-anticodon pattern, there are  also several  rather mysterious points: why some quartets, which encode the same amino acids, are read by one codon and others by two? Why, e.g., the quartet Gly (GGN) is read in some biological species by two anticodons and in other biological species by only one?

Finally let us discuss the case of Table \ref{Table:Ulva}. In this case the  universal pattern remarked above is not generally verified and moreover the value of the most used codon in several cases has a very high value, while other codons are very poorly used. So we may presume that for low value of the GC content or the signs of $c_H$  and $c_V$  and, possibly, of $g_H$  and $g_V$ is different from the one  attributed to the other biological species in the present paper. We are unable to make a more detailed discussion on the basis of only one specimen and we report the Table 21  mainly to highlight the dependance of the constants from the GC content.

\section{Conclusions}

Analyzing the usage frequency inside any codon quadruple of 20 plants, for which the codon-anticodon pattern has been experimentally determined, we have remarked that the hierarchy of the usage frequencies presents an almost ``universal"  behavior and that the hierarchy of codon usage frequencies of the conjugate quadruple is the conjugate one. In order to look for a general explication of this fact we have written the codon usage probability as the sum of two contributions: the first dominant term depends  on the codon-anticodon interaction through an unknown function, assumed decreasing in the value of the interaction, and is an almost ``universal" one in the sense specified in Section 2; the second term is a local one, i.e. depends on the biological species and, may be, on the considered gene. 

First of all we have introduced a model based on the Watson-Crick codon-anticodon
 recognition pattern using the spin-spin codon-anticodon interaction in the crystal basis model. A very symmetric scheme appears, and the results obtained are in satisfactory agreement with the experimental for the whole hierarchy.  In particular for three quadruples: GUN, UUN and AUN, the theoretical results disagree with the observed ones, mainly with a difference in the third and and the fourth  positions between the predicted and the observed hierarchy.
 
Then we have introduced the real codon-anticodon recognition pattern, adding to spin-spin interactions a further z-spin interaction necessary to resolve a degeneracy now present in the previous interaction. From the choice of the signs, with constraints on the parameters, we are able to account for the observed data, except for the complete hierarchy of the quadruple AUN, with a loss of symmetry and some tuning on the values of the parameters of the two interactions.  

So we conclude that the parametrization eq.(\ref{eq:P}) may  be considered as an enough satisfactory approximation. 
Our model strongly suggests that the codon usage bias reflects the following feature:  the most used codon is the best recognized by the anticodon, i.e.  the one which minimize an appropriate function of the codon-anticodon interaction eq.(\ref{eq:T-j}) and so on for the whole rank. Of course it is intriguing to try to determine, from general principles, the form of the function  $F(K_N)$.

  Let us note that we have not at all considered a possible contribution to the ``codon bias" due to ``anticodon bias", for example induced by the abundance of a specific anticodon. This effect may explain the difference between the patten of the codon usage frequencies, in some biological species, and the general pattern.
 
 If we had considered only  the two largest codon usage frequencies, with the ``mirror"  symmetry between conjugate quadruples, we would have had the correct correspondence between the computed and the observed codon usage frequencies hierarchy in  251 cases over the 280 considered ones  in
  the Watson-Crick codon-anticodon recognition pattern, with the signs  of $c_H$ and $c_V$ given in Table \ref{Table:rank-WC} and without any constraint on their values, and  in  265 cases  in the real codon-anticodon recognition pattern given in Table \ref{Table:pat-c-a}, with $g_H > 0$ and $g_V > 0$ and the signs of $c_H$ and $c_V$ and the constraints given in Table \ref{Table:Inter}.
 
  Still  the deep question remains: what does this pattern exhibit? Do the values of the parameters explain the pattern or there is some more deep reason for the appearance of the observed pattern? This is the still big opened question! 

 In closing, our analysis confirms that the codon–anticodon interaction plays a relevant role in the codon usage bias and that the description of this interaction in the crystal basis model is able to take into account some features of the codon usage frequencies. 
 
 In particular, our spin-spin interaction seems to better describe the simple Watson-Crick recognition pattern, while, in the presence of the wobble-mechanism, the interaction has to be modified with the introduction of the z-spin term and with some strong constraints on the constants.
 
 The generally satisfactory agreement of our results with data suggests that it may be interesting to perform further and detailed  analysis taking into account the characteristic of the considered biological species. 
 
 Let us remark that, from our approach, assuming the ``universality" of the values of $c_H$, $c_V$, $g_H$ and  $g_V$, one may derive the anticodon used by a gene of a biological species from the knowledge of the most used codon.

 \section{Appendix 1}
 
 We report the codon usage frequencies computed from the data reported in the caption of each Table, in bold the largest value of the codon usage frequency in the quadruple.
 
\begin{table}[H]
%\scriptsize
\begin{tabular}{|c||c|c|c|c||c|c|c|c|}
 \hline 
        Quadruple & $P_U$ & $P_A$ & $P_C$ & $P_G$ & $P_U$ & $P_A$ & $P_C$ & $P_G$ \\ \hline
        Ala-GCN &{\bf 0,432} & 0,281 & 0,167 & 0,120 &{\bf 0,453} & 0,296 & 0,156 &  0,095 \\ \hline
        Arg-CGN & 0,359 &{\bf 0,374} &  0,128 & 0,139 & 0,362 & {\bf 0,383} &  0,099 & 0,137 \\ \hline
        Pro-CCN &{\bf 0,361} & 0,288 & 0,208 & 0,143 &{\bf 0,385} & 0,295 & 0,196 &   0,124 \\ \hline
        Gly-GGN & 0,303 &{\bf 0,404} &   0,107 & 0,186 & 0,342 &{\bf 0,414} &   0,083 & 0,161 \\ \hline
        Leu-CUN &{\bf 0,442} & 0,286 &  0,131 & 0,141 &{\bf 0,429} & 0,289 & 0,147 &  0,135 \\ \hline
        GAN & 0,355 &{\bf 0,416} &  0,089 & 0,140 & 0,356 &{\bf 0,416} & 0,086 & 0,142 \\ \hline
        Val-GUN & 0,366 &{\bf 0,377} &   0,124 & 0,133 & 0,357 &{\bf 0,371} &  0,135 & 0,136 \\ \hline
        CAN & 0,306 &{\bf 0,462} & 0,093 & 0,139 & 0,322 & {\bf 0,436} & 0,094 & 0,148 \\ \hline
        Ser-UCN &{\bf 0,383} & 0,261 & 0,225 & 0,131 &{\bf 0,375} & 0,287 & 0,221 &  0,117 \\ \hline
        AGN & 0,355 &{\bf 0,413} & 0,097 & 0,135 & 0,348 &{\bf 0,430} &  0,087 & 0,135 \\ \hline
        UUN &{\bf 0,342} & 0,294 &   0,170 & 0,194 &{\bf 0,322} & 0,279 & 0,201 & 0,198 \\ \hline
        AAN & 0,360 &{\bf 0,395} &   0,107 & 0,138 & 0,374 & {\bf 0,383} & 0,106 & 0,137 \\ \hline
        Thr-ACN &{\bf 0,407} & 0,293 & 0,187 &   0,113 &{\bf 0,384} & 0,318 & 0,184 &  0,114 \\ \hline
        AUN &{\bf 0,384} & 0,234 &  0,162 & 0,220 &{\bf 0,365} & 0,251 & 0,170 & 0,214 \\ \hline
\end{tabular}
\centering \caption{ 
Values of the codon usage probabilities for the different quadruples of the biological species: on le left, {\it Salvia mitiorrhiza}, computed from data of Table 3 of  \cite{Salvia} and, on the right, {\it Cocos nucifera}, computed from data of Table 2 of \cite{Coconut}.}
 \label{Table:Sal-Coc}
 %\footnotesize
 
\bigskip
\bigskip
\bigskip

\begin{tabular}{|c||c|c|c|c|||c|c|c|c|}
  \hline
        Quadruple &  $P_U$ & $P_A$ & $P_C$ & $P_G$ & $P_U$ & $P_A$ & $P_C$ & $P_G$ \\ \hline
        Ala-GCN &{
\bf 0,391} & 0,242 & 0,225 &  0,142 &{
\bf 0,410} & 0,296 & 0,172 &   0,122 \\ \hline
        Arg-CGN & 0,317 &{
\bf 0,325} &  0,169 & 0,189 & 0,341 &{
\bf 0,383} &   0,131 & 0,145 \\ \hline
        Pro-CCN & 0,281 &{
\bf 0,376} & 0,217 &   0,126 &{
\bf 0,356} & 0,252 & 0,225 & 0,167 \\ \hline
        Gly-GGN & 0,296 &{
\bf 0,373} &  0,146 & 0,185 & 0,317 &{
\bf 0,365} &  0,121 & 0,197 \\ \hline
        Leu-CUN &{
\bf 0,427} &   0,162 & 0,230 & 0,181 &{
\bf 0,444} & 0,276 &  0,131 & 0,149 \\ \hline
        GAN & 0,353 &{
\bf 0,404} &  0,098 & 0,145 & 0,342 &{
\bf 0,429} &  0,086 & 0,143 \\ \hline
        Val-GUN &{\bf 0,347} & 0,316 & 0,177 &  0,160 & 0,365 &{
\bf 0,372} &   0,130 & 0,133 \\ \hline
        CAN & 0,358 &{\bf 0,407} & 0,096 & 0,139 & 0,301 &{
\bf 0,450} &   0,102 & 0,147 \\ \hline
        Ser-UCN &{
\bf 0,359} &  0,123 & 0,339 & 0,179 &{
\bf 0,375} & 0,252 & 0,217 &  0,156 \\ \hline
        AGN & 0,310 &{
\bf 0,387} & 0,078 & 0,225 & 0,272 &{
\bf 0,319} & 0,224 &  0,185 \\ \hline
        UUN &{\bf 0,349} &  0,154 & 0,259 & 0,238 &{
\bf 0,332} & 0,290 & 0,175 & 0,203 \\ \hline
        AAN &{\bf 0,357} & 0,340 &   0,132 & 0,171 & 0,342 &{
\bf 0,401} & 0,116 & 0,141 \\ \hline
        Thr-ACN &{
\bf 0,336} & 0,210 & 0,288 &   0,166 &{
\bf 0,384} & 0,303 & 0,194 &  0,119 \\ \hline
        AUN &{
\bf 0,353} & 0,245 & 0,221 & 0,181 &{
\bf 0,396} & 0,237 &   0,159 & 0,208 \\ \hline\end{tabular}
\centering \caption{Values of the codon usage probabilities for the different quadruples of the biological species: on the left, {\it  Artemisia selengensis}, computed from data of Table 4 of \cite{Artemi} and, on the right, {\it Utricularia reniformis}, computed from data of Table 2 of \cite{Utric}.}
 \label{Table:Art-Utr}
 \end{table}
 
 \begin{table}[H]
% \scriptsize
\begin{tabular}{|c||c|c|c|c||c|c|c|c|}
   \hline
        Quadruple &  $P_U$ & $P_A$ & $P_C$ & $P_G$ & $P_U$ & $P_A$ & $P_C$ & $P_G$ \\ \hline
        Ala-GCN &{
\bf 0,446} & 0,289 & 0,157 & 0,108 &{
\bf 0,426} & 0,285 & 0,165 & 0,124 \\ \hline
        Arg-CGN &{
\bf 0,380} & 0,379 & 0,108 & 0,133 &{
\bf 0,392} & 0,383 & 0,096 & 0,129 \\ \hline
        Pro-CCN &{
\bf 0,375} & 0,298 & 0,177 & 0,150 &{
\bf 0,379} & 0,296 & 0,194 & 0,131 \\ \hline
        Gly-GGN & 0,322 &{
\bf 0,394} & 0,115 & 0,169 & 0,337 &{
\bf 0,399} & 0,100 & 0,164 \\ \hline
        Leu-CUN &{
\bf 0,445} & 0,280 & 0,141 & 0,134 &{
\bf 0,419} & 0,285 & 0,151 & 0,145 \\ \hline
        GAN & 0,351 &{
\bf 0,409} & 0,090 & 0,150 & 0,355 &{
\bf 0,408} & 0,087 & 0,150 \\ \hline
        Val-GUN & 0,357 &{
\bf 0,379} & 0,127 & 0,137 & 0,365 &{
\bf 0,379} & 0,113 & 0,143 \\ \hline
        CAN & 0,307 &{
\bf 0,457} & 0,094 & 0,142 & 0,323 &{
\bf 0,431} & 0,105 & 0,141 \\ \hline
        Ser-UCN &{
\bf 0,396} & 0,277 & 0,208 & 0,119 &{
\bf 0,369} & 0,272 & 0,229 & 0,130 \\ \hline
        AGN & 0,333 &{
\bf 0,418} & 0,103 & 0,146 & 0,330 &{
\bf 0,422} & 0,097 & 0,151 \\ \hline
        UUN &{
\bf 0,331} & 0,290 & 0,181 & 0,198 &{
\bf 0,321} & 0,289 & 0,186 & 0,204 \\ \hline
        AAN & 0,371 &{
\bf 0,388} & 0,106 & 0,135 & 0,370 &{
\bf 0,376} & 0,112 & 0,142 \\ \hline
        Thr-ACN &{
\bf 0,402} & 0,308 & 0,188 & 0,102 &{
\bf 0,396} & 0,307 & 0,185 & 0,112 \\ \hline
        AUN &{
\bf 0,380} & 0,245 & 0,157 & 0,218 &{
\bf 0,376} & 0,243 & 0,159 & 0,222 \\ \hline\end{tabular}
\centering \caption{Values of the codon usage probabilities for the different quadruples of the  biological species: on the left, {\it  Cynara cardunculus}, computed from data of Table S2 of \cite{Cynara} and, on the right,{\it Aconitum pseudolaeve}, computed from data of Table S3 of \cite{Acon}.}
 \label{Table:Cyn-Aco}
 
\bigskip
\bigskip
\bigskip

\begin{tabular}{|c||c|c|c|c||c|c|c|c|}
 \hline
        Quadruple &  $P_U$ & $P_A$ & $P_C$ & $P_G$ & $P_U$ & $P_A$ & $P_C$ & $P_G$ \\ \hline
        Ala-GCN &{
\bf 0,474} & 0,273 & 0,148 & 0,105 &{
\bf 0,449} & 0,290 & 0,155 & 0,105 \\ \hline
        Arg-CGN & 0,367 &{
\bf 0,384} & 0,126 & 0,123 &{
\bf 0,386} & 0,375 & 0,113 & 0,126 \\ \hline
        Pro-CCN &{
\bf 0,405} & 0,282 & 0,186 & 0,127 &{
\bf 0,392} & 0,292 & 0,177 & 0,139 \\ \hline
        Gly-GGN & 0,341 &{
\bf 0,406} & 0,095 & 0,158 & 0,335 &{
\bf 0,386} & 0,112 & 0,167 \\ \hline
        Leu-CUN &{
\bf 0,451} & 0,287 & 0,138 & 0,124 &{
\bf 0,443} & 0,289 & 0,145 & 0,123 \\ \hline
        GAN & 0,338 &{
\bf 0,453} & 0,080 & 0,129 & 0,356 &{
\bf 0,405} & 0,090 & 0,149 \\ \hline
        Val-GUN &{
\bf 0,382} & 0,359 & 0,120 & 0,139 & 0,358 &{
\bf 0,372} & 0,133 & 0,137 \\ \hline
        CAN & 0,293 &{
\bf 0,485} & 0,095 & 0,127 & 0,309 &{
\bf 0,454} & 0,100 & 0,137 \\ \hline
        Ser-UCN &{
\bf 0,399} & 0,270 & 0,200 & 0,131 &{
\bf 0,409} & 0,267 & 0,209 & 0,115 \\ \hline
        AGN & 0,374 &{
\bf 0,395} & 0,099 & 0,132 & 0,345 &{
\bf 0,400} & 0,103 & 0,152 \\ \hline
        UUN &{
\bf 0,362} & 0,322 & 0,153 & 0,163 &{
\bf 0,322} & 0,291 & 0,184 & 0,203 \\ \hline
        AAN & 0,356 &{
\bf 0,423} & 0,107 & 0,113 & 0,373 &{
\bf 0,376} & 0,118 & 0,133 \\ \hline
        Thr-ACN &{
\bf 0,405} & 0,316 & 0,182 & 0,097 &{
\bf 0,414} & 0,302 & 0,184 & 0,100 \\ \hline
        AUN &{
\bf 0,407} & 0,251 & 0,136 & 0,206 &{
\bf 0,376} & 0,237 & 0,159 & 0,228 \\ \hline
\end{tabular}
\centering \caption{Values of the codon usage probabilities for the different quadruples of the biological species: on the left,  {\it Arabidopsis thaliana}, computed from Table 2 of \cite{Arab} and, on the right, {\it Ageratina adenophora}, computed from Table 3 of \cite{Ager}.}
 \label{Table:Ara-Age}

 \end{table}
 
 \begin{table}[H]
%\scriptsize
\begin{tabular}{|c||c|c|c|c||c|c|c|c|}
 \hline
        Quadruple &  $P_U$ & $P_A$ & $P_C$ & $P_G$ & $P_U$ & $P_A$ & $P_C$ & $P_G$ \\ \hline
        Ala-GCN &{
\bf 0,451} & 0,283 & 0,163 & 0,103 &{
\bf 0,458} & 0,284 & 0,154 & 0,104 \\ \hline
        Arg-CGN & 0,361 &{
\bf 0,411} & 0,107 & 0,121 &{
\bf 0,396} & 0,378 & 0,096 & 0,130 \\ \hline
        Pro-CCN &{
\bf 0,365} & 0,268 & 0,224 & 0,143 &{
\bf 0,393} & 0,294 & 0,198 & 0,115 \\ \hline
        Gly-GGN & 0,338 &{
\bf 0,399} & 0,091 & 0,172 & 0,350 &{
\bf 0,403} & 0,087 & 0,160 \\ \hline
        Leu-CUN &{
\bf 0,443} & 0,290 & 0,134 & 0,133 &{
\bf 0,430} & 0,286 & 0,147 & 0,137 \\ \hline
        GAN & 0,362 &{
\bf 0,415} & 0,080 & 0,143 & 0,345 &{
\bf 0,435} & 0,085 & 0,135 \\ \hline
        Val-GUN & 0,367 &{
\bf 0,392} & 0,107 & 0,134 & 0,359 &{
\bf 0,379} & 0,124 & 0,138 \\ \hline
        CAN & 0,301 &{
\bf 0,478} & 0,089 & 0,132 & 0,311 &{
\bf 0,448} & 0,094 & 0,147 \\ \hline
        Ser-UCN &{
\bf 0,392} & 0,272 & 0,201 & 0,135 &{
\bf 0,377} & 0,281 & 0,219 & 0,123 \\ \hline
        AGN & 0,361 &{
\bf 0,399} & 0,094 & 0,146 & 0,352 &{
\bf 0,434} & 0,082 & 0,132 \\ \hline
        UUN &{
\bf 0,347} & 0,297 & 0,170 & 0,186 &{
\bf 0,330} & 0,290 & 0,182 & 0,198 \\ \hline
        AAN & 0,354 &{
\bf 0,412} & 0,104 & 0,130 & 0,380 &{
\bf 0,391} & 0,099 & 0,130 \\ \hline
        Thr-ACN &{
\bf 0,404} & 0,308 & 0,188 & 0,100 &{
\bf 0,389} & 0,320 & 0,187 & 0,104 \\ \hline
        AUN &{
\bf 0,394} & 0,258 & 0,147 & 0,201 &{
\bf 0,377} & 0,246 & 0,163 & 0,214 \\ \hline\end{tabular}
\centering \caption{Values of the codon usage probabilities for the different quadruples of the biological species: on the left, {\it Lotus japonicus},  computed from Table 2 of \cite{Lotus} and, on the right, {\it Phoenix dactylifera L.},  computed from Table 1 of \cite{Phoenix}.}
 \label{Table:Lot-Pho}

\bigskip
\bigskip
\bigskip 

\begin{tabular}{|c||c|c|c|c||c|c|c|c|}
 \hline
        Quadruple &  $P_U$ & $P_A$ & $P_C$ & $P_G$ & $P_U$ & $P_A$ & $P_C$ & $P_G$ \\ \hline
        Ala-GCN &{
\bf 0,432} & 0,295 & 0,148 & 0,125 &{
\bf 0,426} & 0,265 & 0,173 & 0,136 \\ \hline
        Arg-CGN &{
\bf 0,379} & 0,345 & 0,144 & 0,133 & 0,334 &{
\bf 0,394} & 0,136 & 0,136 \\ \hline
        Pro-CCN &{
\bf 0,398} & 0,268 & 0,215 & 0,119 &{
\bf 0,362} & 0,279 & 0,223 & 0,136 \\ \hline
        Gly-GGN & 0,322 &{
\bf 0,380} & 0,105 & 0,193 & 0,300 &{
\bf 0,381} & 0,106 & 0,213 \\ \hline
        Leu-CUN &{
\bf 0,439} & 0,296 & 0,153 & 0,112 &{
\bf 0,413} & 0,278 & 0,155 & 0,154 \\ \hline
        GAN & 0,319 &{
\bf 0,437} & 0,091 & 0,153 & 0,340 &{
\bf 0,420} & 0,090 & 0,150 \\ \hline
        Val-GUN &{
\bf 0,377} & 0,370 & 0,117 & 0,136 & 0,364 &{
\bf 0,370} & 0,125 & 0,142 \\ \hline
        CAN & 0,303 &{
\bf 0,450} & 0,103 & 0,144 & 0,293 &{
\bf 0,446} & 0,119 & 0,142 \\ \hline
        Ser-UCN &{
\bf 0,361} & 0,239 & 0,287 & 0,113 &{
\bf 0,327} &{
\bf 0,327} & 0,203 & 0,143 \\ \hline
        AGN & 0,333 &{
\bf 0,410} & 0,114 & 0,143 & 0,324 &{
\bf 0,400} & 0,119 & 0,157 \\ \hline
        UUN &{
\bf 0,326} & 0,315 & 0,181 & 0,178 &{
\bf 0,329} & 0,280 & 0,192 & 0,198 \\ \hline
        AAN & 0,314 &{
\bf 0,408} & 0,120 & 0,158 & 0,360 &{
\bf 0,384} & 0,116 & 0,140 \\ \hline
        Thr-ACN &{
\bf 0,421} & 0,272 & 0,192 & 0,115 &{
\bf 0,387} & 0,291 & 0,198 & 0,124 \\ \hline
        AUN &{
\bf 0,386} & 0,228 & 0,152 & 0,235 &{
\bf 0,380} & 0,230 & 0,163 & 0,227 \\ \hline\end{tabular}
\centering \caption{Values of the codon usage probabilities for the different quadruples of the biological species: on the left, {\it Oryza minuta},  computed from Table 5 of \cite{Oryza} and, on the right, {\it Citrus aurantiifolia},  computed from Table S3 of \cite{Citrus}.}
 \label{Table:Ory-Cit}
 \end{table}

 \begin{table}[H]
 %\scriptsize
\begin{tabular}{|c||c|c|c|c||c|c|c|c|}
  \hline
        Quadruple &  $P_U$ & $P_A$ & $P_C$ & $P_G$ & $P_U$ & $P_A$ & $P_C$ & $P_G$ \\ \hline
        Ala-GCN &{
\bf 0,455} & 0,316 & 0,148 & 0,081 &{
\bf 0,456} & 0,286 & 0,151 & 0,107 \\ \hline
        Arg-CGN &{
\bf 0,403} & 0,377 & 0,095 & 0,125 & 0,378 &{
\bf 0,392} & 0,115 & 0,115 \\ \hline
        Pro-CCN &{
\bf 0,406} & 0,298 & 0,186 & 0,110 &{
\bf 0,384} & 0,293 & 0,180 & 0,143 \\ \hline
        Gly-GGN & 0,348 &{
\bf 0,412} & 0,081 & 0,159 & 0,332 &{
\bf 0,411} & 0,102 & 0,155 \\ \hline
        Leu-CUN &{
\bf 0,443} & 0,298 & 0,141 & 0,118 &{
\bf 0,445} & 0,282 & 0,139 & 0,134 \\ \hline
        GAN & 0,344 &{
\bf 0,443} & 0,075 & 0,138 & 0,350 &{
\bf 0,430} & 0,083 & 0,137 \\ \hline
        Val-GUN & 0,364 &{
\bf 0,390} & 0,111 & 0,135 & 0,360 &{
\bf 0,391} & 0,117 & 0,132 \\ \hline
        CAN & 0,332 &{
\bf 0,452} & 0,082 & 0,134 & 0,312 &{
\bf 0,461} & 0,094 & 0,133 \\ \hline
        Ser-UCN &{
\bf 0,391} & 0,270 & 0,220 & 0,119 &{
\bf 0,382} & 0,281 & 0,220 & 0,117 \\ \hline
        AGN & 0,353 &{
\bf 0,426} & 0,083 & 0,138 & 0,329 &{
\bf 0,414} & 0,107 & 0,150 \\ \hline
        UUN &{
\bf 0,330} & 0,304 & 0,176 & 0,190 &{
\bf 0,338} & 0,300 & 0,171 & 0,191 \\ \hline
        AAN & 0,356 &{
\bf 0,402} & 0,104 & 0,138 & 0,368 &{
\bf 0,387} & 0,112 & 0,133 \\ \hline
        Thr-ACN &{
\bf 0,394} & 0,317 & 0,174 & 0,115 &{
\bf 0,415} & 0,293 & 0,181 & 0,111 \\ \hline
        AUN &{
\bf 0,389} & 0,259 & 0,144 & 0,208 &{
\bf 0,380} & 0,254 & 0,152 & 0,214 \\ \hline\end{tabular}
\centering \caption{Values of the codon usage probabilities for the different quadruples of the biological species: on the left, {\it Amonum compactum},  computed from Table 3  of \cite{Amonum} and, on the right, {\it Morus cathayana},  computed from Table S1 of \cite{Morus}.}
 \label{Table:Amo-Mor}

\bigskip
\bigskip
\bigskip
 
\begin{tabular}{|c||c|c|c|c||c|c|c|c|}
  \hline
        Quadruple &  $P_U$ & $P_A$ & $P_C$ & $P_G$ & $P_U$ & $P_A$ & $P_C$ & $P_G$ \\ \hline
        Ala-GCN &{
\bf 0,454} & 0,270 & 0,153 & 0,123 &{
\bf 0,487} & 0,341 & 0,163 & 0,009 \\ \hline
        Arg-CGN & 0,367 &{
\bf  0,391} & 0,118 & 0,124 &{
\bf 0,393} & 0,386 & 0,099 & 0,122 \\ \hline
        Pro-CCN &{
\bf 0,376} & 0,280 & 0,203 & 0,141 &{
\bf 0,411} & 0,290 & 0,182 & 0,117 \\ \hline
        Gly-GGN & 0,326 &{
\bf 0,381} & 0,118 & 0,175 & 0,350 &{
\bf 0,413} & 0,077 & 0,160 \\ \hline
        Leu-CUN &{
\bf 0,430} & 0,274 & 0,148 & 0,148 &{
\bf 0,455} & 0,289 & 0,137 & 0,119 \\ \hline
        GAN & 0,343 &{
\bf 0,440} & 0,082 & 0,135 & 0,346 &{
\bf 0,425} & 0,075 & 0,154 \\ \hline
        Val-GUN & 0,356 &{
\bf 0,386} & 0,119 & 0,139 & 0,334 &{
\bf 0,367} & 0,109 & 0,190 \\ \hline
        CAN & 0,308 &{
\bf 0,468} & 0,093 & 0,131 & 0,317 &{
\bf 0,458} & 0,089 & 0,136 \\ \hline
        Ser-UCN &{
\bf 0,378} & 0,271 & 0,227 & 0,124 &{
\bf 0,392} & 0,272 & 0,215 & 0,121 \\ \hline
        AGN & 0,341 &{
\bf 0,413} & 0,103 & 0,143 & 0,347 &{
\bf 0,426} & 0,086 & 0,141 \\ \hline
        UUN &{
\bf 0,335} & 0,310 & 0,167 & 0,188 &{
\bf 0,331} & 0,299 & 0,177 & 0,193 \\ \hline
        AAN & 0,368 &{
\bf 0,394} & 0,107 & 0,131 & 0,360 &{
\bf 0,395} & 0,106 & 0,139 \\ \hline
        Thr-ACN &{
\bf 0,402} & 0,301 & 0,184 & 0,113 &{
\bf 0,393} & 0,318 & 0,172 & 0,117 \\ \hline
        AUN &{
\bf 0,387} & 0,257 & 0,149 & 0,207 &{
\bf 0,387} & 0,263 & 0,142 & 0,208 \\ \hline\end{tabular}
\centering \caption{Values of the codon usage probabilities for the different quadruples of the biological species: on the left, {\it Castanopsis sclerophylla},  computed from Table S2  of \cite{Cast} and, on the right, {\it Zingiber montanum},  computed from Table S4 of \cite{Zing}.}
 \label{Table:Cast-Zing}
 \end{table}

\begin{table}[H]
%\scriptsize
\begin{tabular}{|c||c|c|c|c||c|c|c|c|}
   \hline
        Quadruple &  $P_U$ & $P_A$ & $P_C$ & $P_G$ & $P_U$ & $P_A$ & $P_C$ & $P_G$ \\ \hline
        Ala-GCN &{
\bf 0,462} & 0,283 & 0,157 & 0,098 &{
\bf 0,464} & 0,279 & 0,150 & 0,107 \\ \hline
        Arg-CGN &{
\bf  0,398} & 0,381 & 0,092 & 0,128 & 0,372 &{
\bf  0,379} & 0,116 & 0,133 \\ \hline
        Pro-CCN &{
\bf 0,393} & 0,295 & 0,206 & 0,106 &{
\bf 0,394} & 0,291 & 0,183 & 0,132 \\ \hline
        Gly-GGN & 0,345 &{
\bf 0,411} & 0,098 & 0,146 & 0,325 &{
\bf 0,415} & 0,095 & 0,165 \\ \hline
        Leu-CUN &{
\bf 0,422} & 0,294 & 0,143 & 0,141 &{
\bf 0,437} & 0,294 & 0,140 & 0,129 \\ \hline
        GAN & 0,339 &{
\bf 0,433} & 0,083 & 0,145 & 0,346 &{
\bf 0,434} & 0,083 & 0,137 \\ \hline
        Val-GUN & 0,355 &{
\bf 0,383} & 0,128 & 0,134 &{
\bf  0,373} & 0,359 & 0,127 & 0,141 \\ \hline
        CAN & 0,312 &{
\bf 0,461} & 0,090 & 0,137 & 0,293 &{
\bf 0,474} & 0,096 & 0,137 \\ \hline
        Ser-UCN &{
\bf 0,392} & 0,262 & 0,222 & 0,124 & 0,307 &{
\bf 0,319} & 0,223 & 0,151 \\ \hline
        AGN & 0,367 &{
\bf 0,415} & 0,083 & 0,135 & 0,353 &{
\bf 0,403} & 0,107 & 0,137 \\ \hline
        UUN &{
\bf 0,328} & 0,294 & 0,183 & 0,195 &{
\bf 0,355} & 0,307 & 0,167 & 0,171 \\ \hline
        AAN & 0,369 &{
\bf 0,394} & 0,109 & 0,128 & 0,356 &{
\bf 0,412} & 0,107 & 0,125 \\ \hline
        Thr-ACN &{
\bf 0,384} & 0,310 & 0,199 & 0,107 &{
\bf 0,404} & 0,311 & 0,179 & 0,106 \\ \hline
        AUN &{
\bf 0,385} & 0,250 & 0,151 & 0,214 &{
\bf 0,393} & 0,251 & 0,149 & 0,207 \\ \hline
\end{tabular}
\centering \caption{Values of the codon usage probabilities for the different quadruples of the biological species: on the left, {\it Musa acuminata},  computed from Table 2  of \cite{Musa} and, on the right, {\it Brassica oleracea},  computed from Table 6 of \cite{Bras}.}
 \label{Table:Musa-Bras}

\bigskip
\bigskip
\bigskip

\begin{tabular}{|c||c|c|c|c||c|c|c|c|}
   \hline
        Quadruple &  $P_U$ & $P_A$ & $P_C$ & $P_G$ & $P_U$ & $P_A$ & $P_C$ & $P_G$ \\ \hline
        Ala-GCN &{
\bf 0,435} & 0,298 & 0,159 & 0,108 &{
\bf 0,469} & 0,297 & 0,142 & 0,092 \\ \hline
        Arg-CGN &{
\bf  0,392} & 0,350 & 0,124 & 0,134 & 0,381 &{
\bf  0,405} & 0,102 & 0,112 \\ \hline
        Pro-CCN &{
\bf 0,389} & 0,296 & 0,190 & 0,125 &{
\bf 0,380} & 0,315 & 0,190 & 0,115 \\ \hline
        Gly-GGN & 0,330 &{
\bf 0,398} & 0,100 & 0,172 & 0,341 &{
\bf 0,403} & 0,092 & 0,164 \\ \hline
        Leu-CUN &{
\bf 0,447} & 0,275 & 0,152 & 0,126 &{
\bf 0,451} & 0,292 & 0,132 & 0,125 \\ \hline
        GAN & 0,344 &{
\bf 0,404} & 0,090 & 0,162 & 0,347 &{
\bf 0,437} & 0,085 & 0,131 \\ \hline
        Val-GUN & 0,357 &{
\bf 0,367} & 0,141 & 0,135 &{
\bf  0,376} &{
\bf  0,376} & 0,111 & 0,137 \\ \hline
        CAN & 0,299 &{
\bf 0,442} & 0,104 & 0,155 & 0,314 &{
\bf 0,477} & 0,084 & 0,125 \\ \hline
        Ser-UCN &{
\bf 0,374} & 0,283 & 0,229 & 0,114 &{
\bf 0,391} & 0,292 & 0,197 & 0,120 \\ \hline
        AGN & 0,340 &{
\bf 0,434} & 0,084 & 0,142 & 0,347 &{
\bf 0,416} & 0,103 & 0,134 \\ \hline
        UUN &{
\bf 0,341} & 0,292 & 0,178 & 0,189 &{
\bf 0,356} & 0,301 & 0,163 & 0,180 \\ \hline
        AAN &{
\bf 0,396} & 0,371 & 0,096 & 0,136 &{
\bf 0,398} & 0,395 & 0,097 & 0,110 \\ \hline
        Thr-ACN &{
\bf 0,393} & 0,299 & 0,176 & 0,132 &{
\bf 0,426} & 0,314 & 0,156 & 0,104 \\ \hline
        AUN &{
\bf 0,377} & 0,264 & 0,154 & 0,205 &{
\bf 0,405} & 0,286 & 0,136 & 0,173 \\ \hline\end{tabular}
\centering \caption{Values of the codon usage probabilities for the different quadruples of the biological species: on the left, {\it Epipremnum aureum},  computed from Table 4  of \cite{Epip} and, on the right, {\it Glycine soja},  computed from Table 4 of \cite{Glyc}.}
 \label{Table:Epip-Glyc}
\end{table}

\begin{table}[H]
%\scriptsize
\begin{tabular}{|c||c|c|c|c||c|c|c|c|}
  \hline
        Quadruple & $P_U$ & $P_A$ & $P_C$ & $P_G$ \\ \hline
        Ala-GCN &{
\bf 0,523} & 0,400 & 0,037 & 0,040 \\ \hline
        Arg-CGN &{
\bf 0,734} & 0,184 & 0,062 & 0,020 \\ \hline
        Pro-CCN & 0,428 &{
\bf 0,456} & 0,054 & 0,062 \\ \hline
        Gly-GGN &{
\bf 0,611} & 0,258 & 0,067 & 0,064  \\ \hline
        Leu-CUN &{
\bf 0,515} & 0,345 & 0,075 & 0,065  \\ \hline
        GAN & 0,388 &{
\bf 0,506} & 0,055 & 0,051  \\ \hline
        Val-GUN &{
\bf 0,583} & 0,366 & 0,020 & 0,031 \\ \hline
        CAN & 0,253 &{
\bf 0,608} & 0,079 & 0,060  \\ \hline
        Ser-UCN & 0,407 &{
\bf 0,517} & 0,027 & 0,049 \\ \hline
        AGN &{
\bf 0,520} & 0,358 & 0,071 & 0,051  \\ \hline
        UUN & 0,354 &{
\bf 0,562} & 0,055 & 0,028 \\ \hline
        AAN & 0,380 &{
\bf 0,391} & 0,099 & 0,130  \\ \hline
        Thr-ACN & 0,392 &{
\bf 0,508} & 0,041 & 0,059  \\ \hline
        AUN &{
\bf 0,557} & 0,230 & 0,052 & 0,161  \\ \hline\end{tabular}
\centering \caption{Values of the codon usage probabilities for the different quadruples of the biological species{\it Ulva flexuosa},  computed from Table S4 of \cite{Ulva}.}
 \label{Table:Ulva}
 \end{table}

  \section{Appendix 2}

 In the crystal basis model, nucleotides are classified in the fundamental representation (denoted by $(1/2;1/2)$ and of dimensionality 4) of the Quantum Group, see e.g. \cite{K},  $U_{q \to 0} (SU_L(2) \otimes SU_V(2))$ and the eigenvalues of the generators $J_{H,3}$ and $J_{V,3}$  on the four bases are chosen as (see \cite{FSS98} and \cite{SS17}):
\begin{eqnarray}
&SU(2)_{H}& \nonumber \\
C \equiv (+ 1/2,+1/2) &\qquad\longleftrightarrow\qquad& U \equiv (-1/2,+1/2) 
\nonumber \\
 \Big. SU(2)_{V} \updownarrow && \updownarrow SU(2)_{V} \\
\Big. G \equiv (+1/2,-1/2) &\qquad\leftrightarrow\qquad& A \equiv (-1/2,-1/2) 
\nonumber \\
&SU(2)_{H}& \nonumber
\label{eq:basis}
\end{eqnarray}

The subscripts $H$ (:= horizontal) and $V$ (:= vertical) 
 are used to distinguish the two $U_{q \to 0}SU(2)$, their initials being inspired from the above scheme. Consequently  codons and anticodons, which are obtained by the the  three-fold Kronecker product of the fundamental representation, are classified in representations of $U_{q \to 0} (SU_L(2) \otimes SU_V(2))$.

Let us define the 2-rows column vectors  and the 2-columns row vectors  basis
\be
\left( \begin{array}{c}
1 \\
0  \end{array} \right)
\qquad
\left( \begin{array}{c}
0 \\
1  \end{array} \right),
\qquad
\qquad
\left( \begin{array}{cc}
1 &
0  \end{array} \right)
\qquad
\left( \begin{array}{cc}
0 &
1  \end{array} \right)
\label{eq:(2r-2c)}
\ee
which can be seen, respectively, as the 2-dimensional representation of $U_{q \to 0}SU_V(2)$, acting on the right of the column vector, and of  $U_{q \to 0}(SU_H(2))$, acting, after transposition, on the left of row vector.
In this basis the generator of $U_{q \to 0}(SU(2))$ are

\be
J_+ = \left( \begin{array}{cc}
0 & 1 \\
0  & 0 \end{array} \right)
\qquad
J_- = \left( \begin{array}{cc}
0 & 0  \\
1 & 0  \end{array} \right)
\qquad
J_3 =\left( \begin{array}{cc}
1/2 & 0 \\
0 & - 1/2  \end{array} \right)
\ee

Let us consider the outer product of the 2-rows column vectors  and the 2-columns row vectors, written in  eq.(\ref{eq:(2r-2c)}),
which gives a $2 \times 2$ matrix, $\mathbb{B}$, which can be considered as the 4-dimensional fundamental irreducible representation  $(1/2;1/2)$ of $U_{q \to 0} (SU_L(2) \otimes SU_V(2))$. 

This $2 \times 2$ matrix, describing the 4 bases (nucleotides), can be written, in compact notation referring to eq.(\ref{eq:basis}), as
\be
\centering
 \mathbb{B} =  \left[\begin{matrix} C &  U   \\  G & A  \end{matrix} \right]
\ee
 Then we consider the $4 \times 4$ matrix, Kronecker or direct product of two $\mathbb{B}$

\be
\centering
\mathbb{B}_1 \otimes \mathbb{B}_2 =  \left[\begin{matrix}\mathbb{B}_1 C &  \mathbb{B}_1 U   \\  \mathbb{B}_1 G & \mathbb{B}_1 A  \end{matrix} \right] = \left[\begin{array}{cc:cc}
CC & UC  & CU & UU  \\ GC  & AC & GU & AU  \\
\hline
CG & (UG) & CA & (UA) \\  GG & AG  & GA & AA  \\
\end{array}\right]
\label{eq:array1}
\ee

 $\mathbb{B}_1 \otimes \mathbb{B}_2$ is the matrix representing the 16-dim reducible representation $(1/2;1/2) \otimes (1/2;1/2)$, describing the di-nucleotides. $\mathbb{B}_1 = \mathbb{B}_2$ but as we are concerned with not commutative quantities, e.g. $CA \neq AC$, we write the lower labels 1 and 2 to specify the procedure used to perform the direct product.

\noindent In brackets the di-nucleotides which form quadruples containing Stop codons and  which will not  be considered.

 In \cite{SS12}   we have described the codon-anticodon interaction by means of an operator
  ${\mathbf T}$ given by
  \be
{\mathbf T} = 8 c_H \,\vec{ J_H^c} \cdot  \vec{J_H^a} + 8  c_V \, \vec{J_V^c} \cdot \vec{ J_V ^a}
 \label{eq:T}
\ee
 which has to be computed between the 
 ``states" identifying the codon and anticodon in the  ``crystal basis model",  where:

\begin{itemize}
\item $J_H^c ,  J_V^c$  (respectively $J_H^a ,  J_V^a$) are the labels of  ${\mathcal U_{q \to 0}}(su(2)_H \oplus su(2)_V)$  specifying the state describing the codon $XZN $ (respectively the anticodon $N'Z_cX_c$ reading the codon  $XYZ$).

\item  $\vec{ J_{\alpha}^c} \cdot  \vec{J_{\alpha}^a}$ ($\alpha = H, V$) should be read as 
\be
\vec{ J_{\alpha}^c} \cdot  \vec{J_{\alpha}^a} =
\frac{1}{2}\left\{\left(\vec{J_{\alpha}^c}  \oplus \vec{J_{\alpha}}^a \right)^2 - ( \vec{J_{\alpha}^c})^2 -  ( \vec{J_{\alpha}^a})^2 \right\}
\ee
and   $\vec{J_{\alpha}^c}  \oplus  \vec{J_{\alpha}^a} \equiv \vec{J_{\alpha}^T}$  stands for the irreducible representation which the codon-anticodon state under consideration belongs to, the tensor product of   $\vec{J_{\alpha}^c}$ and $\vec{J_{\alpha}^a}$ being performed according to the rule of  \cite{K}, choosing the codon as first vector and the anticodon as second vector.  Note that 
$ \vec{J_{\alpha}}^2$ should be read as the Casimir operator whose eigenvalues are given by $
J_{\alpha}(J_{\alpha} +1 )$.
\end{itemize}
 
Let us recall that the expression $ < N'Z_cX_c| \mathbf{T}|XZN> $ has to be read as
 \bea
&& T_{NN'}  \equiv < N'Z_cX_c| \mathbf{T} |XZN>   \equiv  \mathbf{T} \,  \left(|XZN> \otimes \,  | N'Z_cX_c > \right) \nonumber \\
&& = \mathbf{T} \, \left( |J_H^c, J_V^c;J_{H,3}^c, J_{V,3}^c > \otimes \,  |J_H^a, J_V^a;J_{H,3}^a, J_{V,3}^a >  \right)  \nonumber \\
&& =  \lambda \,   \left( |J_H^c, J_V^c;J_{H,3}^c, J_{V,3}^c > \otimes \, |J_H^a, J_V^a;J_{H,3}^a, J_{V,3}^a >  \right) 
\eea
where we have used the correspondence
\bea
|XZN>  & \; \ra & |J_H^c, J_V^c;J_{H,3}^c, J_{V,3}^c > \nonumber \\  
  |N'Z_cX_c >  & \ra  &|J_H^a, J_V^a;J_{H,3}^a, J_{V,3}^a >
\eea
and $\lambda$ is  the eigenvalue of  $\mathbf {T}$  on the state $|J_H^c, J_V^c;J_{H,3}^c, J_{V,3}^c > \otimes \,  |J_H^a, J_V^a;J_{H,3}^a, J_{V,3}^a > $, see  \cite{SS12}  for more details.
 
 \bigskip
 
{\bf Acknowledgements} - The authors sincerely thank the referees for their constructive criticism and their comments, that have improved the clarity of the manuscript.

% \newpage

\end{document}